# Implementation of Security Features in Software Development Phases

1st Ariessa Davaindran Lingham
*School of Computer Science & Engineering,*
*Taylor's University*
Selangor, Malaysia
writingsdystopia@gmail.com

2nd Nelson Tang Kwong Kin
*School of Computer Science & Engineering,*
*Taylor's University*
Selangor, Malaysia
nelsontang0507@gmail.com

3rd Chen Wan Jing
*School of Computer Science & Engineering,*
*Taylor's University*
Selangor, Malaysia
Janecwj18@gmail.com

4th Chong Heng Loong
*School of Computer Science & Engineering,*
*Taylor's University*
Selangor, Malaysia
chonghengloong@sd.taylors.edu.my

5th Fatima-tuz-Zahra
*School of Computer Science & Engineering,*
*Taylor's University*
Selangor, Malaysia
fatemah.tuz.zahra@gmail.com

**Abstract** – Security holds an important role in a software. Most people are not aware of the significance of security in software system and tend to assume that they will be fine without security in their software systems. However, the lack of security features causes to expose all the vulnerabilities possible to the public. This provides opportunities for the attackers to perform dangerous activities to the vulnerable insecure systems. This is the reason why many organizations are reported for being victims of system security attacks. In order to achieve the security requirement, developers must take time to study so that they truly understand the consequences and importance of security. Hence, this paper is written to discuss how secure software development can be performed. To reach the goal of this paper, relevant researches have been reviewed. Multiple case study papers have been studied to find out the answers to how the vulnerabilities are identified, how to eliminate them, when to implement security features, why do we implement them. Finally, the paper is concluded with final remarks on implementation of security features during software development process. It is expected that this paper will be a contribution towards the aforementioned software security domain which is often ignored during practical application.

## 1  Introduction

A secure software is a software where it is designed to protect the contents from being attacked resulting in facing loss [1]. Mouelhi et al. (2015) stated that a software without vulnerabilities can be said to be highly secure as the possibility of the software being attacked is usually based on the number of vulnerabilities exist in the software. This can be further explained with even one existing vulnerability is insecure as it opens up the chance for attackers to perform attacks towards the software. With the evolution and implementation of insecure technologies like Internet of Things (IoT) [2] the chances of attacks have grown much more in recent years. Therefore, it is necessary to address these issues as well as critically review, analyze and update software development processes [3] in terms of integration of security features.

Many developers focus on the functional requirements while developing a software as that is the first impression left for users when they use the software. However, they are neglecting the consequences of not managing non-functional requirements like security properly, which will lead them to deep regrets and huge loss later on. Some do aware of the importance of implementing security in a software system but did an inefficient way of doing it. Developers often look for vulnerabilities at the end of the project development just to realize that there are many vulnerabilities and had to make changes abruptly. To solve that, it is worth emphasizing that security implementations during the process of software development is the ideal way to save time and work efficiently.

With that being said, there are measures that can assist developers in planning the procedure of security implementations. The secure software development lifecycle (SDLC) is a very useful framework to synchronize their security knowledge with developing the software from the beginning to the end [4]. There are phases in SDLC which leads developers to better development, designing and maintenance of the software making sure all the functional, non-functional requirements and objectives are achieved [5]. This reduces the risk of making mistakes involving security during development process as well as improving the quality production. Such strategy allows earlier discovery of vulnerabilities therefore, earlier amendment of issues.

There are rules to declare a software has achieved security, which is by achieving Confidentiality, Integrity, and Availability (CIA) [6]. Confidentiality means that information is not disclosed to anyone who are unauthorized [7]. Breach of confidentiality could occur when one's data or information is shared to a third party without the owner's consent [8]. Integrity can be defined as a piece of information is not being altered or changed [7]. Integrity is violated when someone unauthorized manage to get permission to make changes or edit anything related to the data [9]. While Scarfone et al. (2019) says availability is to be able to provide services to authorized users without having interruptions such as system down. System down could sometimes be attackers performing denial of service attack where they flood server with non-ending requests to connect until the server could not respond to each of them and causes the system to stop functioning [10]. Some other attacks that could be launched on vulnerable systems include ransomware [11] and malware attacks where the victim computers, mobiles or smart devices like smartphones [12] are infected with malicious software to launch the attack. Therefore, additional rules and regulations have been proposed and implemented in some cases which may assist in achieving security, including authentication as well as authorization [6].

## 2 Literature Review

The concept of Software Security relies on making the software able to behave in a certain manner when exposed to a malicious entity and or attack. However, software malfunctions often happen without any intentional malicious attacks or targeting [13]. In the past, this concept was not held in high priority, as it was widely believed that a simple security network infrastructure would be enough to stave off attacks by malicious users, however, this has proven not to be enough particularly now when the use of insecure systems is frequent in critical systems and organizations, such as healthcare, transportation [14] and surveillance systems. Hence, it is crucial that given the overwhelming amount of attacks that are faced by high profile companies and governments, the security and technological implementations, physical and digital, should be highly secure in order to reduce the risks of data leakage and thievery [15]. This has formed the SDLC (Software Development Life Cycle) which is a set of development tactics that helps with planning to design secure software. These procedures and techniques should be utilized in all steps of software development and maintenance in order to protect the data and ensure the best outcome. As in most cases, SDLC lifecycles generally have six main phases as shown in Fig. 1. Most planning and cost for errors in each stage vary, hence preparation for each stage is crucial for a smooth sailing software security development process. The phases are discussed in forthcoming sections.

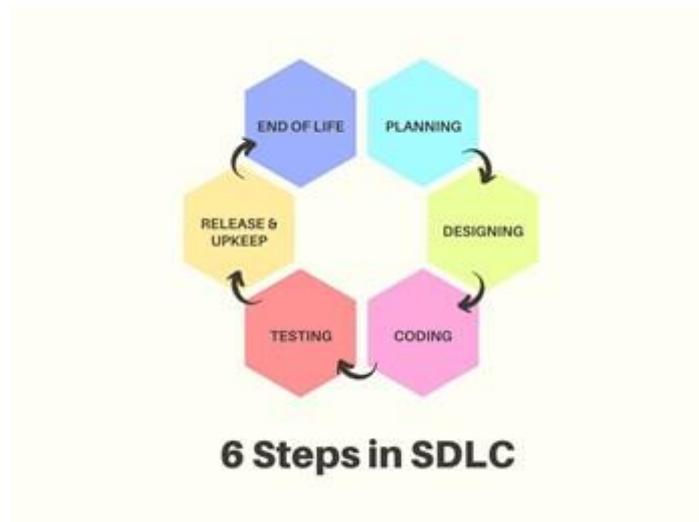

Fig. 1. Steps in software development life cycle

### 2.1 Planning (Requirements Gathering)

Also known as requirements gathering, this is the first stage of the SDLC six step cycle. This includes determining the costs and resources necessary to develop the project in a progressive manner, along with sub-plans and plan B's should anything out of the plan occur.

*A. Identifying SDL and CMMI*

SDL (Security Development Lifecycle) and CMMI (Capability Maturity Model) [16][17] are two components in SDLC that are crucial to properly planning the requirements, risks and costs of the project components. Microsoft SDL [18] components are a list of planning ideologies to follow to create a better software project. As for CMMI , the golden rule is that it stands for the fact that the overall final quality of the product is defined and highly influenced by the materials and processes used to create it. There are 3 major CMMI models, which overlap each other providing more specific focus into Development, Service Management and Acquisition.

- *CMMI - DEV :* This is the development CMMI model, which mainly offers the current time best practices for development, procurement of necessary resources and maintenance of the software, including certain tools to aid companies in enhancing their productivity and processes alongside guidelines for process maturity and capacity.
- *CMMI - SWC :* This is the CMMI model responsible for service management, which provide service provider organizations with tips and guides for developing, providing and maintaining services.
- *CMMI - ACQ :* This is the CMMI model [19] that deals with acquisitions organizations in order to develop and procure best guidelines in order to handle the acquisitions of resources, including goods and or services.

*B. Project Scheduling*

This includes identifying the project scopes in order to prevent any scope creeps from running up on schedule later on in development. Scheduling entails breaking down all necessary portions of the project into smaller sized chunks to work on, determining dependency of the functions and categorizing necessary work by priority on the development list. Resource allocations for the project is usually done through the use of a GANTT chart, which is then converted to a PERT chart after all resource allocations are completed.

*C. Identifying Necessary Security Implementations*

Its good practice to define the types of security requirements necessary for developing the program, which would increase the chances of making a successful application through careful planning and risk assessment. Overall, there are three main types of security requirements [20]:

- *Functional security requirements*

These describe what the system must do, such as functional behavior enforcing security in the application, and hence can directly be tested for any issues. Examples of these include authentication, encryptions, data integrity systems and access controls.

- *Non-functional security requirements*

Non-functional Security Requirements entail what a system must become - such as statements that provide insight into uptime, app flexibility, and usability. Most of these are mainly statements that give the team and client a clear guideline on how to build certain security features of the app.

- *Derived security requirements*

Derived Security Requirements include portions of the functional and non-functional requirements, defining more detailed characteristics of both. An example includes defining the number of guesses a user has for their password verification code before they are put on timer cool down. Most Derived security requirements require the planner to think both as a user, and an attacker to properly lay out the foundation of what sub-requirements are necessary for each requirement.

As such, the project team is then able to evaluate how and how much protection should be incorporated into the software during different stages of development. Most of the requirements are listed during this phase, however with threat models and changing attack trends, industry standard reports can be referenced, such as the IEEE (Institute of Electrical and Electronics Engineers), ISO (International Organization of Standardization) , and NIST (National Institute of Standards and Technology).

*D. Developing Misuse Cases (and Misuse Actor Maps)*

The opposite of a use case would be a misuse case, which is defined as a function that the system must not allow to happen - "… a completed sequence of actions which results in loss for the organization or some specific stakeholder." [21] as quoted by Guttom Sindre and Andreas Opdahl's "Templates for Misuse Case Description" [21]. Generally, misuse cases are very helpful in checking for security requirements, and help with understanding and distinguishing between different types of malicious attackers and entities.

**2.2 Designing Phase**

This process works from the initial on-paper ideas and focuses on the "how to" portion of the project, including designing the system's architecture [22] and how each method and function should be executed and enforced during the requirement phase. Besides the system architecture, the project team will also produce an SRS document (Software Requirements Specification) which is a description of how every feature will be implemented, alongside framework layouts and other scenarios, additionally including options to select other resources such as third party components to speed up development.

*A. Developing Encryption Implementations*

Encryptions are crucial in any secure application and program, hence a solid encryption strategy must be developed. Although encryption can be created into a feature, to implement it in the overall design of the application as a fundamental implementation would save much cost. There are several components to decide and implement encryption into the application:

*1) Defining the items to be protected*

Most data should be encrypted as it bypasses through the internet and the application itself, especially in private networks. In addition, it's advised that organizations create specific rules and guidelines for data protection.

*2) Defining the cryptographic mechanisms to be used*

For better data security, it is wise to use only industry approved and standard encryption libraries, algorithms, key lengths and cipher styles.

*3) Defining the type of encryption key and certificate suitable*

Ensuring the adequate encryption keys and data management systems are incredibly important, and work hand in hand with data encryption as a whole. In order to ensure safety of the cipher keys and certificates, it's advised to create a system in which monitoring and handling the keys is easily traceable in the case of any mishaps. There are many encryption-based solutions to avoid security attacks proposed by researchers, such as pre-encryption algorithm to detect crypto-ransomware [23] attack and lightweight encryption strategies [24] for IoT systems. However they should be tested for reliable implementation

B. *Developing Threat Models*

Threat models are chosen based on the cases of misuse identified in the planning stage. Threat models help calculate risks and manage resources efficiently in different attack scenarios, including the eventual planning and development of a more secure program as it influences the security architecture, testing and code quality control. Microsoft has come out with the STRIDE model (Fig. 2.) as a sample threat modelling system, aiming to classify risks by different categories of their attacks onto different points in the system. DREAD, too is an acronym that utilizes a score based system to assess risks, by calculating the ratings.

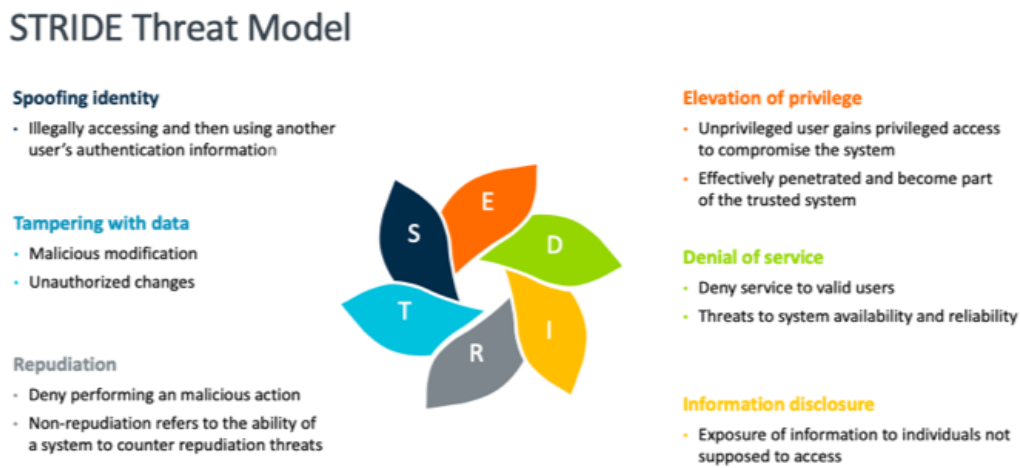

Fig. 2. Example of STRIDE threat model [25]

**2.3    Coding and Implementation Phase**

After all requirements and system design have been concluded, coding and development begins, in which programmers will follow the specifications listed in the previous two phases to program the application. This includes building the base code, running code reviews and quality control tests during development, and creating stable builds for testing. Secure code [26] is written, then automated tools for checking review the static code, before a final manual review takes place.

A. *Writing Secure Code*

When coding functions, developers should be well informed about vulnerabilities in the code that can be faced during implementation. A threat model approach, such as designed in the Designing phase, is necessary to predict and plan for serious risks, and take extra precautions in doing so. A major aim of designing a secure application is reducing bug numbers in the code during implementation itself, which would reduce the amount of unintended vulnerabilities and lower the cost for later debugging.

B.  *Handling Errors*

The main aspect of the program security would be to deal with any unexpected attacks or errors in a systematic manner, sending error codes and or recovering / resetting as the situation calls for it. As such, error handling should be implemented into logging methods and errors should be directed to log file administrators as automated feedback for quick action should anything go catastrophically. However, what exactly the errors and error codes displayed entail should not be shown to the customers and users of the website, to prevent any potential attackers from finding the weak point in the system to exploit. Error messages should be clear and could have hyperlinks or direct instructions to direct the user to help pages or other resources to instruct customers on how to proceed [27].

C.  *Using Safe Functions*

Since 2002, Microsoft has isolated and began researching the safety of common C and C++ functions that were prone to buffer overflow attacks and coding errors. As such, the Safe String Functions [28][29] were created to help replace certain faulty functions in these languages. Although C and C++ are a prime example of the threats of using unsafe functions without proper and thorough checking of how the data is travelling, several other languages have functions that may be difficult to implement safely. Hence in the project, developers must be ensured to have sufficient knowledge on certain unsafe functions and their safer counterparts, besides learning proper industry security coding standards. Certain third party tools such as certain IDEs can be used to locate and analyze the usage of certain perilous functions. Besides this, linting, a basic static code analyzing system and static analysis have general frameworks and guidelines for defining the usage of different functions during building [30].

D.  *Static Code Analysis*

Static analysis of code ensures that the code to be tested is of the highest quality possible before any revisions. With the right analysis tool, development is sped up as well, by finding out certain errors and weaknesses in the code as below:

- Programming errors
- Security weaknesses in code
- Standard coding violations

Static analyzers are very good at finding issues such as buffer overflow opportunities, null pointers and possible memory leaks as well. As Static analyzers are low cost, simple and efficient to use, these help test the code's integrity by helping the software to be more secure besides piling up on all the latest security implementations.

## 2.4  Testing Phase

Testing the software is a highly important part of the SDLC cycle [31], which helps weed out usage function errors, cases of misuse and abuse, identifying potential threats, as well as danger models during the test preparation.

A.  *Creating a Testing Plan*

The goal of testing is to report, monitor, sort out and keep testing the software until determined to fit to the standard in the SRS's quality requirements [32]. Many teams may use RTM (Requirements Traceability Matrix) documents to track the requirements of the project and log their interactions.

*B. Fuzzing*

In the words of H.D. Moore, Chief Architect of Metasploit, fuzzing is "… the process of sending intentionally invalid data to a product in the hopes of triggering an error condition or fault." [33]. As such, fuzzing is a key method in order to simulate different types of error inputs into the system to find any hidden errors and bugs that have been overlooked during manual code reviews and automated tool scouring. Fuzzing runs an endless amount of different checks against the system, with one of the higher resources used being CPU energy, but the results of generating better code quality and awareness through such testing is one of its major advantages.

*C. Dynamic Testing*

Dynamic based testing schemes focus on the ways used to gauge software quality, validating the usefulness of the software through the executions of the software under testing conditions which comprise of real or simulated inputs in various formats under different operating conditions to give a real look at how the software has a possibility to recover, respond and continue under attack or in the presence of an unusual input [34]. DAST (Dynamic Application Security Testing) is the study of the code execution is what makes it possible to adjust the quality level of the software, and see whether the requirements for such are met to the SRS standards. It focuses on the complete operation, unlike SAST (Static Application Security Testing) which is based on the syntax.

*D. Penetration Testing*

Tests may detect the greatest amount of bugs and flaws with tools to evaluate the program or device to find out any potential weak spots there may be, evaluating the system's behavior, and how it communicates on the network to other programs or devices as part of the commands it carries out. There are several types of penetration testing for an all rounded report [35]:

- *External testing:*

External penetration tests try targeting company assets that pass through the internet, such as organization websites, emails and DNS servers. The goal of this testing would be to attempt gaining access to and extracting data.

- *Internal testing:*

During internal testing, testers given access to the program behind any protective security implementations on the outside (i.e. firewalls) simulate attacks by a malicious inside party, such as a staff, etc. These can include rogue employees to stolen credentials.

- *Targeted testing:*

Targeted testing entails the tester and organization security personnel to work together in order to ensure that the security personnel have a view into how hackers operate on the system, enabling more efficient training of security staff. This is a training exercise that gives the security team real time feedback from how a hacker would act, and ways to counter it.

- *Blind testing:*

In a blind test, the tester is only given the name of the organization to target, and hence they have to figure out ways to get in. Security personnel can take this opportunity to see how an actual assault works.

- *Double blind testing:*

In a double blind testing scenario, the security personnel do not have any idea or knowledge about the attack to be simulated, which represents attempted breaches in the real world. As such, this provides valuable training and knowledge resources to the organization and developers of the program, in order to make their application as secure as necessary.

## 2.5 Release and Upkeep (Maintenance)

Once reaching this stage of the SDLC, it's safe to assume that all the prior testing scenarios have been passed, and the application is safe for deployment. Before release however, the product must be reviewed and perform a quality control check before going live. Maintenance keeps the app up to its most secure state with new patches to stay updated with the current industry security requirements.

*A. Initializing Security Reviews*

After testing, a software analysis is in store, which has the purpose to identify any vulnerabilities that have not been rooted out earlier. To prevent loss of too many resources spent on security assessment, the reviews should be well planned. The program's general condition, functionality and standards are all put through a quality control test, undergoing various tests and checks including for bugs and errors that have been encountered previously to see if they will appear again.

*B. Initializing Incident Response Personnel*

An Incident Response Team includes people assigned to implement the response plan, generally consisting of IT staff collecting, preserving and analyzing data for correlations and other data related to incidents. The Incident Response plan is crucial in most cases to ensure that there is no utter chaos that follows after a breach has occurred, and that the organization can move forward smoothly while dealing with the issue at large. CSIRTs (Computer Security Incident Response Teams) are expected to be familiar with the policies, guidelines, and other related knowledge pertaining to vulnerability solving, analysis as well as guidelines to reporting any incidents. They should also be able to direct the software engineer responsible for fixing the issue through the problems and possible solutions as well [36]. Fig. 3. shows six steps which can be used to build proactive and effective incident response plan for cyber attack-related occurrences.

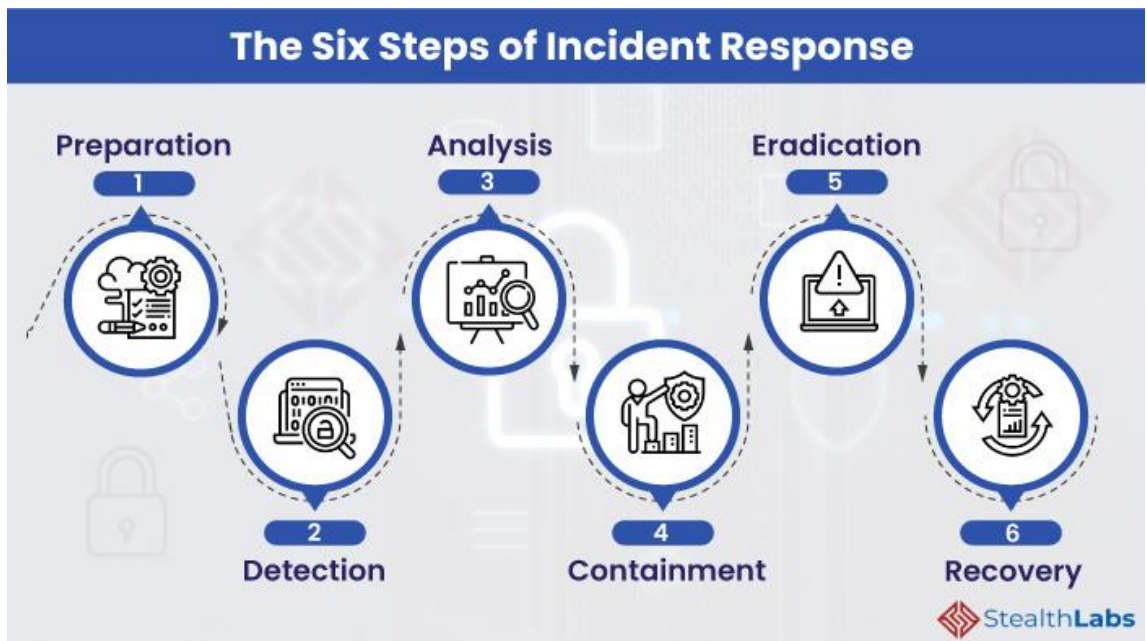

Fig.3. Steps to develop proactive incident response plan [37]

## 2.6 End of Life

This is the point when the software is no longer in top shape and the developers stop producing support patches or updates for the application. End of life [38] regulations can be used on most applications holding sensitive data such as user credentials.

*A. Creating Data Disposal Policies*

At the end of the application, valuable data must be properly disposed of, including customer data and encryption keys. If done correctly, these policies will help safeguard information from ill-meaning parties by destroying them and prevent data breaches. Provided it does not interfere with certain legal requirements, such as data retention and protection policies, all user data should be purged from the system if it is no longer required for the operation of the application.

*B. Data Retention*

Different governments have different policies as to retention of user data. Ensuring the company policies for data retention match the rules set by the law helps eliminate possibilities of being fined for non-compliance. An example to follow would be the ISO 15489 standard [39], which states that all retention conditions and treatment procedures should be created in such a way that it will protect against loss, damage, misuse, and unauthorized access, amongst others [40]. Besides this, organizations are expected to have policies and guidelines on conversions or movement of data and records from system to system.

## 3 Methodology

Due to the ongoing CMCO, our options for research methods were quite limited so for the literature review, we were only able to do a research on online articles related to security in software development. The research is done through examining, comparing and understanding the concerns and practices highlighted in the articles to form only tangible conclusions for our research. Some examples of the articles we researched were:
1) "Security Risks in the Software Development Lifecycle" which explains the strategies and practices involved in implementing risk management throughout the SDLC as well as the analysis of current risk practices(best practices, pitfalls and mitigation) [41].
2) "A Preventive Secure Software Development Model for a Software Factory" which is a case study comparing reactive security with preventive security approach in development models [42].

With this research method, we get information that are not as bias and opinion-based as, say a questionnaire, because it doesn't depend on the past experience of a small respondent group. Many of the articles we researched are based on past experiments, real data and the writers themselves have most probably also done their own extensive research. For example, in the 2[nd] article above, the author did an experiment that found that vulnerabilities could be reduced by 68.42% by using development models with preventive security approach [42]. We also get a wider understanding of security concerns and practices from online articles because it offers more thorough justification and comparison. The writers of these articles have spent time understanding the problem and putting together the solution they think is best. If we did a survey, the practices given by respondents may only be specific to their situation. Although we were not able to perform other research methods, one of the article we reviewed included detailed information about an interview with a local software engineer. The interviewee was asked about his job scope, the development model used, security guidelines followed and the challenges implementing the security guideline. The interviewee's team had an application security team to oversee the security aspect of the development, the interviewee mentioned challenges communicating with them which ended up with a clash of security and functional requirements.

## 4  Discussion

There are several software development strategies depending on the nature of system required for a particular domain of application. Numerous surveys have been performed to measure their effectiveness, efficiency as well as reliability in terms of security, constructivism and suitability for practical implementation [43]. Several aspects have been uncovered through those reviews and analysis. Similarly, through the research carried out in this paper it is found that it is quite common to see software development companies disregarding security matters in the early stages of the software development lifecycle (SDLC). By doing so, developers are bound to introduce vulnerabilities into the software and as the development progresses, the earlier vulnerabilities get more complicated to fix as new layers of code gets added. As a result, the company will have to spend much more resources to enhance software security in the future. Almost all the articles are suggesting that security aspect be taken more if not as seriously as the functional requirements. What this means is that when we plan and design the functional requirements, we also need to consider the possible security hole that we might be creating. The articles are suggesting to employ use case and misuse case, use case is a classic way of defining how users would interact with the system, while misuse case defines how users can exploit a function to attack the system. This would require developers to be rather proficient with cybersecurity first which is why many articles are also suggesting security training.

Security is also an important thing to keep in mind when implementing the software, especially in today's so called modern age where wireless sensor networks [44], IoT and similar resource-constrained technologies are insecurely implemented in sensitive environments. We can design the software to be extremely secure but some issues may not be apparent until we actually start the implementation. For example, some vulnerabilities like buffer overflow are produced only in the implementation phase because the developer either forgets to sanitize user input or have allocated insufficient memory for certain variables. The articles suggest to mitigate these vulnerabilities by defining a secured coding standards/guideline and require developers to follow it strictly. Other than that, companies wanting to implement secure development process will often refer to the OWASP's [45] top 10 list of most critical security risk. However, this list should only be treated as a starting point instead of the final goal as a test conducted by Micro Focus found that 49% of discovered vulnerabilities were not in the list. In addition, two common vulnerabilities; the cross-site request forgery(CSRF) and unvalidated redirects, were taken out the list in 2017.

## 5  Solution

Some software engineers lack the security training and it becomes an issue in the software development security domain. It is difficult for security experts to persuade developers to think about security of their code. Some solutions are suggested to deal with the challenge and issue.

*A.  Integration of Security Review and Testing into Implementation Phase*

Periodic security testing or review should be implemented on the code so that instant and further feedback can be received by the developers to review their code's quality and improve the security, An automated spell-checker tool can be used to implement the test as the code will be scrutinized by a security expert to do pair programming with agile methodology. Two experts will work on the code together and the roles of driver and observer will be switched regularly. Pair programming is considered as one of the effective approaches since there are nighty-five percent of the developers are more confident to code by using pair programming while there are fifteen percent lesser defects in the code with collaborative pairs [46].

*B.  Mentorship of Security*

A group of security experts should be hired by companies to mentor developers to learn about security concepts and knowledges since they have no much experience and knowledge. Incorporating lessons can be provided to the developers as feedback can be provided at any time during inspection and review of the code. A strong foundation of software security can be built by the developers with the help of the mentors which have much experience for leading them to the right direction. Therefore, the wrong information or bad coding practice can be prevented by the developers as these cases might be happened during learning throughself- study [47].

*C. Utilization of Automated Security Tools*

The process to secure software can be automated by automated security tools such as vulnerability scanner as constant monitoring on software is provided by the tools to identify vulnerabilities while an automated report will be generated. The workload of security engineers and software engineers will be reduced with the help of the tools and the vulnerabilities are documented in the report generated automatically. The security team is allowed by the tools for enforcement of security policies to be applied on the development environment. Thus, the security team is allowed for locating and fixing vulnerabilities or bugs as soon as possible [48].

*D. Education of Software Security and Secure Software Design*

Some developers lack the security training and this might cause a bad integration of security in software project while both software and security requirements are clashed. Some developers do not think of their software from the perspective of the attackers because they lack security training. Agile development will have some vulnerabilities for emphasizing flexibility and speed that the developers may ignore the security part of the software. Therefore, some potential threats may be faced by the software so the developers must have a security training to consider security part from the perspective of attackers [48].

## 6 Conclusion

In conclusion, cyber-attacks are increasing and the attackers are improving their ways to target the systems that they want to attack. Therefore sufficient knowledge and techniques are required to secure software systems. In this modern era, secure software development frameworks and practices are used by some organizations to secure their systems in order to prevent data breaches and other security attacks. Thus, confidentiality of the data can be ensured. Effectiveness of some security practices is extremely high to reinforce software security and the security practices are considered as one of the standards for the companies. Some security frameworks and benchmarks such as Microsoft SDL are used to help decision making of security model implementation for the companies. The security frameworks also help in providing knowledge about the safety level of an organization's current security practices. There are some challenges that the relevant stakeholder may face during implementation of secure software development, but the companies can make an effort to improve the environment for secure software development in order to handle the challenges. Lastly, there is no perfect and everlasting solution to secure software development since cyber-attacks are becoming more complex and evolving. Therefore, the developers should always update their knowledge and skills of software security and use robust security tools to ensure the security of the developed system.